\input epsf
\input amssym
\newfam\scrfam
\batchmode\font\tenscr=rsfs10 \errorstopmode
\ifx\tenscr\nullfont
        \message{rsfs script font not available. Replacing with calligraphic.}
        \def\scr{\cal}
\else   
        \font\sevenscr=rsfs7
        \font\fivescr=rsfs5
        \skewchar\tenscr='177 \skewchar\sevenscr='177 \skewchar\fivescr='177
        \textfont\scrfam=\tenscr \scriptfont\scrfam=\sevenscr
        \scriptscriptfont\scrfam=\fivescr
        \def\scr{\fam\scrfam}
        \def\cal{\scr}
\fi
\catcode`\@=11
\newfam\frakfam
\batchmode\font\tenfrak=eufm10 \errorstopmode
\ifx\tenfrak\nullfont
        \message{eufm font not available. Replacing with italic.}
        
\else
	
	\font\sevenfrak=eufm7 \font\fivefrak=eufm5
	\textfont\frakfam=\tenfrak
	\scriptfont\frakfam=\sevenfrak \scriptscriptfont\frakfam=\fivefrak
	
\fi
\catcode`\@=\active
\newfam\msbfam
\batchmode\font\twelvemsb=msbm10 scaled\magstep1 \errorstopmode
\ifx\twelvemsb\nullfont\def\Bbb{\bf}

	\message{Blackboard bold not available. Replacing with boldface.}
\else   \catcode`\@=11
        \font\tenmsb=msbm10 \font\sevenmsb=msbm7 \font\fivemsb=msbm5
        \textfont\msbfam=\tenmsb
        \scriptfont\msbfam=\sevenmsb \scriptscriptfont\msbfam=\fivemsb
        \def\Bbb{\relax\expandafter\Bbb@}
        \def\Bbb@#1{{\Bbb@@{#1}}}
        \def\Bbb@@#1{\fam\msbfam\relax#1}
        \catcode`\@=\active

\fi
        \font\eightrm=cmr8              \def\xrm{\eightrm}
        \font\eightbf=cmbx8             \def\xbf{\eightbf}
        \font\eightit=cmti10 at 8pt     \def\xit{\eightit}
        
                     
        \font\eightcp=cmcsc8
        \font\eighti=cmmi8              \def\xold{\eighti}
        \font\eightib=cmmib8             \def\xbold{\eightib}
        \font\teni=cmmi10               \def\old{\teni}
        \font\tencp=cmcsc10

        \font\twelvecp=cmcsc10 scaled\magstep1

	 at10pt	
	\font\twelvehelvbold=phvb at12pt
	 at14pt
	\font\sixteenhelvbold=phvb at16pt

\def\noblackbox{\overfullrule=0pt}
\noblackbox

\newtoks\headtext
\headline={\ifnum\pageno=1\hfill\else
	\ifodd\pageno{\eightcp\the\headtext}{ }\dotfill{ }{\old\folio}
	\else{\old\folio}{ }\dotfill{ }{\eightcp\the\headtext}\fi
	\fi}
\def\makeheadline{\vbox to 0pt{\vss\noindent\the\headline\break
\hbox to\hsize{\hfill}}
        \vskip2\baselineskip}
\newcount\infootnote
\infootnote=0
\def\foot#1#2{\infootnote=1
\footnote{${}^{#1}$}{\vtop{\baselineskip=.75\baselineskip
\advance\hsize by -\parindent\noindent{\xrm #2}}}\infootnote=0$\,$}
\newcount\refcount
\refcount=1
\newwrite\refwrite
\def\oldsize{\ifnum\infootnote=1\xold\else\old\fi}
\def\ref#1#2{
	\def#1{{{\oldsize\the\refcount}}\ifnum\the\refcount=1\immediate\openout\refwrite=\jobname.refs\fi\immediate\write\refwrite{\item{[{\xold\the\refcount}]} 
	#2\hfill\par\vskip-2pt}\xdef#1{{\noexpand\oldsize\the\refcount}}\global\advance\refcount by 1}
	}
\def\refout{\catcode`\@=11
        \xrm\immediate\closeout\refwrite
        \vskip2\baselineskip
        {\noindent\twelvecp References}\hfill\vskip\baselineskip
        \baselineskip=.75\baselineskip
        \input\jobname.refs
        \baselineskip=4\baselineskip \divide\baselineskip by 3
        \catcode`\@=\active\rm}

\def\hepth#1{\href{http://xxx.lanl.gov/abs/hep-th/#1}{hep-th/{\xold#1}}}

\def\arxiv#1#2{\href{http://arxiv.org/abs/#1.#2}{arXiv:{\xold#1}.{\xold#2}}}
\def\jhep#1#2#3#4{\href{http://jhep.sissa.it/stdsearch?paper=#2\%28#3\%29#4}{J. High Energy Phys. {\xbold #1#2} ({\xold#3}) {\xold#4}}}

\def\ATMP#1#2#3{Adv. Theor. Math. Phys. {\xbold#1} ({\xold#2}) {\xold#3}}

\def\CQG#1#2#3{Class. Quantum Grav. {\xbold#1} ({\xold#2}) {\xold#3}}

\def\JHEP{\jhep}

\def\NPB#1#2#3{Nucl. Phys. {\xbf B}{\xbold#1} ({\xold#2}) {\xold#3}}

\def\PLB#1#2#3{Phys. Lett. {\xbf B}{\xbold#1} ({\xold#2}) {\xold#3}}

\newcount\sectioncount
\sectioncount=0
\def\section#1#2{\global\eqcount=0
	\global\subsectioncount=0
        \global\advance\sectioncount by 1
	\ifnum\sectioncount>1
	        \vskip2\baselineskip
	\fi
\line{\twelvecp\the\sectioncount. #2\hfill}
       \vskip.5\baselineskip\noindent
        \xdef#1{{\old\the\sectioncount}}}
\newcount\subsectioncount
\def\subsection#1#2{\global\advance\subsectioncount by 1
	\vskip.75\baselineskip\noindent
\line{\tencp\the\sectioncount.\the\subsectioncount. #2\hfill}
	\vskip.5\baselineskip\noindent
	\xdef#1{{\old\the\sectioncount}.{\old\the\subsectioncount}}}
\def\immediatesubsection#1#2{\global\advance\subsectioncount by 1
\vskip-\baselineskip\noindent
\line{\tencp\the\sectioncount.\the\subsectioncount. #2\hfill}
	\vskip.5\baselineskip\noindent
	\xdef#1{{\old\the\sectioncount}.{\old\the\subsectioncount}}}
\newcount\appendixcount
\appendixcount=0
\def\appendix#1{\global\eqcount=0
        \global\advance\appendixcount by 1
        \vskip2\baselineskip\noindent
        \ifnum\the\appendixcount=1
        \hbox{\twelvecp Appendix A: #1\hfill}\vskip\baselineskip\noindent\fi
    \ifnum\the\appendixcount=2
        \hbox{\twelvecp Appendix B: #1\hfill}\vskip\baselineskip\noindent\fi
    \ifnum\the\appendixcount=3
        \hbox{\twelvecp Appendix C: #1\hfill}\vskip\baselineskip\noindent\fi}
\def\acknowledgements{\vskip2\baselineskip\noindent
        \underbar{\it Acknowledgements:}\ }
\newcount\eqcount
\eqcount=0
\def\Eqn#1{\global\advance\eqcount by 1
\ifnum\the\sectioncount=0
	\xdef#1{{\old\the\eqcount}}
	\eqno({\oldstyle\the\eqcount})
\else
        \ifnum\the\appendixcount=0
	        \xdef#1{{\old\the\sectioncount}.{\old\the\eqcount}}
                \eqno({\oldstyle\the\sectioncount}.{\oldstyle\the\eqcount})\fi
        \ifnum\the\appendixcount=1
	        \xdef#1{{\oldstyle A}.{\old\the\eqcount}}
                \eqno({\oldstyle A}.{\oldstyle\the\eqcount})\fi
        \ifnum\the\appendixcount=2
	        \xdef#1{{\oldstyle B}.{\old\the\eqcount}}
                \eqno({\oldstyle B}.{\oldstyle\the\eqcount})\fi
        \ifnum\the\appendixcount=3
	        \xdef#1{{\oldstyle C}.{\old\the\eqcount}}
                \eqno({\oldstyle C}.{\oldstyle\the\eqcount})\fi
\fi}
\def\eqn{\global\advance\eqcount by 1
\ifnum\the\sectioncount=0
	\eqno({\oldstyle\the\eqcount})
\else
        \ifnum\the\appendixcount=0
                \eqno({\oldstyle\the\sectioncount}.{\oldstyle\the\eqcount})\fi
        \ifnum\the\appendixcount=1
                \eqno({\oldstyle A}.{\oldstyle\the\eqcount})\fi
        \ifnum\the\appendixcount=2
                \eqno({\oldstyle B}.{\oldstyle\the\eqcount})\fi
        \ifnum\the\appendixcount=3
                \eqno({\oldstyle C}.{\oldstyle\the\eqcount})\fi
\fi}
\def\multi{\global\advance\eqcount by 1}
\def\multieq#1#2{\xdef#1{{\old\the\eqcount#2}}
        \eqno{({\oldstyle\the\eqcount#2})}}
\newtoks\url
\def\Href#1#2{\catcode`\#=12\url={#1}\catcode`\#=\active#2}
\def\href#1#2{{#2}}

\parskip=3.5pt plus .3pt minus .3pt
\baselineskip=14pt plus .1pt minus .05pt
\lineskip=.5pt plus .05pt minus .05pt
\lineskiplimit=.5pt
\abovedisplayskip=18pt plus 4pt minus 2pt
\belowdisplayskip=\abovedisplayskip
\hsize=14cm
\vsize=19cm
\hoffset=1.5cm
\voffset=1.8cm
\frenchspacing
\footline={}
\raggedbottom

\def\ss{\scriptstyle}
\def\sss{\scriptscriptstyle}
\def\*{\partial}
\def\punkt{\,\,.}
\def\komma{\,\,,}

\def\={\!=\!}
\def\small#1{{\hbox{$#1$}}}

\def\fraction#1{\small{1\over#1}}
\def\fr{\fraction}
\def\Fraction#1#2{\small{#1\over#2}}
\def\Fr{\Fraction}

\def\eg{{\tenit e.g.}}

\def\ie{{\tenit i.e.}}

\def\a{\alpha}
\def\b{\beta}

\def\d{\delta}
\def\e{\varepsilon}
\def\g{\gamma}
\def\l{\lambda}

\def\th{\theta}

\def\ra{\rightarrow}

\def\ra{\rightarrow}
\def\la{\leftarrow}

\def\rarrowover#1{\vtop{\baselineskip=0pt\lineskip=0pt
      \ialign{\hfill##\hfill\cr$\ra$\cr$#1$\cr}}}

\def\larrowover#1{\vtop{\baselineskip=0pt\lineskip=0pt
      \ialign{\hfill##\hfill\cr$\la$\cr$#1$\cr}}}


\def\modprod#1{\raise0pt\vtop{\baselineskip=0pt\lineskip=0pt
      \ialign{\hfill##\hfill\cr$\circ$\cr${\sss #1}$\cr}}}

\def\third{3\raise3pt\hbox{\eightit rd}}
\def\fourth{4\raise3pt\hbox{\eightit th}}

\def\TPsi{\tilde\Psi}
\def\tpsi{\tilde\psi}


\ref\CederwallNilssonTsimpisI{M. Cederwall, B.E.W. Nilsson and D. Tsimpis,
{\xit ``The structure of maximally supersymmetric super-Yang--Mills
theory---constraining higher order corrections''},
\jhep{01}{06}{2001}{034} 
[\hepth{0102009}].}

\ref\CederwallNilssonTsimpisII{M. Cederwall, B.E.W. Nilsson and D. Tsimpis,
{\xit ``D=10 super-Yang--Mills at $\ss O(\a'^2)$''},
\JHEP{01}{07}{2001}{042} [\hepth{0104236}].}

\ref\BerkovitsParticle{N. Berkovits, {\xit ``Covariant quantization of
the superparticle using pure spinors''}, \jhep{01}{09}{2001}{016}
[\hepth{0105050}].}

\ref\SpinorialCohomology{M. Cederwall, B.E.W. Nilsson and D. Tsimpis,
{\xit ``Spinorial cohomology and maximally supersymmetric theories''},
\jhep{02}{02}{2002}{009} [\hepth{0110069}];
M. Cederwall, {\xit ``Superspace methods in string theory,
supergravity and gauge theory''}, Lectures at the XXXVII Winter
School in Theoretical Physics ``New Developments in Fundamental 
Interactions Theories'',  Karpacz, Poland,  Feb. 6-15, 2001, \hepth{0105176}.}

\ref\Movshev{M. Movshev and A. Schwarz, {\xit ``On maximally
supersymmetric Yang--Mills theories''}, \NPB{681}{2004}{324}
[\hepth{0311132}].}

\ref\BerkovitsI{N. Berkovits,
{\xit ``Super-Poincar\'e covariant quantization of the superstring''},
\jhep{00}{04}{2000}{018} [\hepth{0001035}].}

\ref\BerkovitsNonMinimal{N. Berkovits,
{\xit ``Pure spinor formalism as an N=2 topological string''},
\jhep{05}{10}{2005}{089} [\hepth{0509120}].}

\ref\CederwallNilssonSix{M. Cederwall and B.E.W. Nilsson, {\xit ``Pure
spinors and D=6 super-Yang--Mills''}, \arxiv{0801}{1428}.}

\ref\CGNN{M. Cederwall, U. Gran, M. Nielsen and B.E.W. Nilsson,
{\xit ``Manifestly supersymmetric M-theory''},
\JHEP{00}{10}{2000}{041} [\hepth{0007035}];
{\xit ``Generalised 11-dimensional supergravity''}, \hepth{0010042}.
}

\ref\CGNT{M. Cederwall, U. Gran, B.E.W. Nilsson and D. Tsimpis,
{\xit ``Supersymmetric corrections to eleven-dimen\-sional supergravity''},
\jhep{05}{05}{2005}{052} [\hepth{0409107}].}

\ref\NilssonPure{B.E.W.~Nilsson,
{\xit ``Pure spinors as auxiliary fields in the ten-dimensional
supersymmetric Yang--Mills theory''},
\CQG3{1986}{{\xrm L}41}.}

\ref\HowePureI{P.S. Howe, {\xit ``Pure spinor lines in superspace and
ten-dimensional supersymmetric theories''}, \PLB{258}{1991}{141}.}

\ref\HowePureII{P.S. Howe, {\xit ``Pure spinors, function superspaces
and supergravity theories in ten and eleven dimensions''},
\PLB{273}{1991}{90}.} 

\ref\CederwallBLG{M. Cederwall, {\xit ``N=8 superfield formulation of
the Bagger--Lambert--Gustavsson model''}, \jhep{08}{09}{2008}{116}
[\arxiv{0808}{3242}].}

\ref\CederwallABJM{M. Cederwall, {\xit ``Superfield actions for N=8 
and N=6 conformal theories in three dimensions''},
\jhep{08}{10}{2008}{70}
[\arxiv{0808}{3242}].}

\ref\ElevenSG{E. Cremmer, B. Julia and J. Scherk, 
{\xit ``Supergravity theory in eleven-dimensions''},
\PLB{76}{1978}{409}.}

\ref\ElevenSGSuperspace{L. Brink and P. Howe, 
{\xit ``Eleven-dimensional supergravity on the mass-shell in superspace''},
\PLB{91}{1980}{384};
E. Cremmer and S. Ferrara,
{\xit ``Formulation of eleven-dimensional supergravity in superspace''},
\PLB{91}{1980}{61}.}
 
\ref\BatalinVilkovisky{I.A. Batalin and G.I. Vilkovisky, {\xit ``Gauge
algebra and quantization''}, \PLB{102}{1981}{27}.}

\ref\FusterBVReview{A. Fuster, M. Henneaux and A. Maas, {\xit
``BRST-antifield quantization: a short review''}, \hepth{0506098}.}

\ref\BerkovitsMembrane{N. Berkovits,
	{\xit ``Towards covariant quantization of the supermembrane''},
	\JHEP{02}{09}{2002}{051} [\hepth{0201151}].}

\ref\BerkovitsNekrasovMultiloop{N. Berkovits and N. Nekrasov, {\xit
    ``Multiloop superstring amplitudes from non-minimal pure spinor
    formalism''}, \jhep{06}{12}{2006}{029} [\hepth{0609012}].}

\ref\AnguelovaGrassiVanhove{L. Anguelova, P.A. Grassi and P. Vanhove,
  {\xit ``Covariant one-loop amplitudes in D=11''},
  \NPB{702}{2004}{269} [\hepth{0408171}].}

\ref\GrassiVanhove{P.A. Grassi and P. Vanhove, {\xit ``Topological M
    theory from pure spinor formalism''}, \ATMP{9}{2005}{285}
  [\hepth{0411167}].} 

\ref\PureSG{M. Cederwall, {\xit ``Towards a manifestly supersymmetric
    action for D=11 supergravity''}, \arxiv{0912}{1814}, to appear in
  J. High Energy Phys.}


\headtext={M. Cederwall: ``D=11 supergravity with manifest supersymmetry''}

\line{
\epsfxsize=18mm
\epsffile{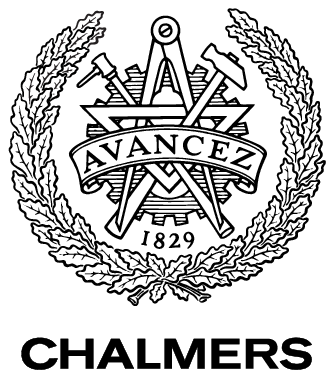}
\hfill}
\vskip-12mm
\line{\hfill G\"oteborg preprint}
\line{\hfill December, {\old2009}}
\line{\hrulefill}

\vfill
\vskip.5cm

\centerline{\sixteenhelvbold
D=11 supergravity} 

\vskip4\parskip

\centerline{\sixteenhelvbold
with manifest supersymmetry} 

\vfill

\centerline{\twelvehelvbold
Martin Cederwall}

\vfill

\centerline{\it Fundamental Physics}
\centerline{\it Chalmers University of Technology}
\centerline{\it SE 412 96 G\"oteborg, Sweden}

\vfill

{\narrower\noindent\underbar{Abstract:} The complete supersymmetric action for
  eleven-dimensional supergravity is presented. The action is
  polynomial in the scalar fermionic pure spinor superfield, and contains only a
  minor modification to the recently proposed three-point coupling.
\smallskip}
\vfill

\font\xxtt=cmtt6

\vtop{\baselineskip=.6\baselineskip\xxtt
\line{\hrulefill}
\catcode`\@=11
\line{email: martin.cederwall@chalmers.se\hfill}
\catcode`\@=\active
}

\eject

\def\l{\lambda}
\def\lb{\bar\lambda}


\section\Introduction{Introduction}Eleven-dimensional 
supergravity [\ElevenSG] is the low-energy limit of the (not yet
defined) M-theory, and hence of a
strong-coupling limit of string theory. 
Having maximal supersymmetry, a traditional superspace description
[\ElevenSGSuperspace] puts the theory on-shell. Recently, a programme
was initiated to formulate eleven-dimensional supergravity with
manifest supersymmetry using a pure spinor superfield. A simple
three-point interaction was proposed [\PureSG].

Pure spinor superfields provide a powerful
tool for formulating supersymmetric field and string theories
[\NilssonPure\hbox to6cm{\phantom{\HowePureI,\HowePureII,\BerkovitsI,
\BerkovitsParticle,\CGNN,\CederwallNilssonTsimpisI,
\CederwallNilssonTsimpisII,\SpinorialCohomology,\Movshev,\CGNT,
\BerkovitsNonMinimal,\CederwallNilssonSix,
\CederwallBLG}\hfill}\hskip-6cm-\CederwallABJM]. 
In models with maximal supersymmetry, the constraint on the ordinary
superfield, which enforces the equations of motion, is encoded in a
cohomological equation of the type $Q\Psi+\ldots=0$, which is the equation of
motion for the pure spinor superfield.
Pure spinor superfield
theory inevitably leads to a Batalin--Vilkovisky (BV) formalism 
[\BatalinVilkovisky,\FusterBVReview].

In the present paper, we will show that the deformation of the free
action represented by the three-point interaction of ref. [\PureSG] is
almost the whole answer. Due to the simple properties of the operators
involved, higher order interactions are essentially absent.
Pure spinor superfield formulations tend to have some remarkable
properties, as an extra bonus in addition to the manifest
supersymmetry. 
The action for $D=10$ super-Yang--Mills is
Chern--Simons-like, and has only a cubic interaction
[\BerkovitsParticle]. 
The conformal
models in $D=3$, whose component actions contain couplings of six
scalar fields, simplify enormously in the pure spinor framework, where
the matter superfields only have a minimal coupling to the
Chern--Simons field [\CederwallBLG,\CederwallABJM]. 
Higher order interactions arise when auxiliary
fields are eliminated (in both cases the fermionic component of the gauge
connection on superspace). This type of simplification is shared by
$D=11$ supergravity, surprisingly to the extent that the action
becomes polynomial.

The organisation of the paper is as follows. In Section {\old2}, we review
the construction of ref. [\PureSG]. The full action is given in
Section {\old3}, where we also expand the action around a
background. Section {\old4} contains conclusions and a discussion, where we
focus on identifying future directions of research.

\section\Review{Eleven-dimensional supergravity with pure spinors}The
relevant pure
spinors in $D=11$ satisfy $(\l\g^a\l)=0$. It has been known for some
time that the cohomology of a scalar fermionic superfield under
the the BRST operator $q=(\l D)$ gives the linearised supergravity
multiplet [\CGNN,\SpinorialCohomology,\BerkovitsMembrane]. 
The fermionic derivatives anticommute to give torsion 
$\{D_\a,D_\b\}=-T_{\a\b}{}^cD_c$, with
$T_{\a\b}{}^c=-2\g_{\a\b}^c$. This holds for any
background satisfying the equations of motion, but for all
purposes in this paper $D_\a$ will be the flat covariant derivative. 
The fermionic 
scalar field $\Psi$ has dimension $-3$ and ghost number $3$, and its lowest
component is the third order ghost for the tensor field. The physical
fields of ghost number 0 sit in the field as $\l^\a\l^\b\l^\g
C_{\a\b\g}(x,\th)$, where $\l^\a$ is the pure spinor and $C_{\a\b\g}$
the lowest-dimensional part of the superspace 3-form $C$. 
There is a natural measure on the pure spinor space,
and it is straightforward to write an action $\int\Psi Q\Psi$ 
giving the linearised 
equations of motion [\BerkovitsMembrane]. 
The integrand has ghost number 7 and dimension
$-6$. We refer to ref. [\PureSG] for details and conventions.

In refs. [\CGNN,\SpinorialCohomology] it was shown that there is also another
field, $\Phi^a$, of dimension $-1$ and ghost number $1$, that contains
the linearised multiplet. This field has the additional gauge symmetry 
$\Phi^a\approx\Phi^a+(\l\g^a\rho)$, and has as its lowest component
the diffeomorphism ghost. The physical fields sit in $\l^\a h_\a{}^a$,
where $h_\a{}^a$ is the linearised supervielbein.

It is necessary, both for having a non-degenerate measure and in order
to write the relevant operators, to work with non-minimal pure
spinors [\BerkovitsNonMinimal]. In addition to the pure spinor $\l$,
one has the pure spinor $\lb$ and the fermionic spinor $r$ which is
pure relative to $\lb$, $(\lb\g^ar)=0$. The non-minimal BRST operator
is $Q=q+s=(\l D)+(r{\*\over\*\lb})$.

\subsection\IntroOne{The three-point coupling}In our recent paper
[\PureSG], we constructed a three-point coupling for
eleven-dimensional supergravity. This was done by constructing the
BRST-invariant operator $R^a$ relating the two fields according to 
$\Phi^a=R^a\Psi$. It takes the form
$$
\eqalign{
R^a&=R_0^a+R_1^a+R_2^a\cr
&=\eta^{-1}(\lb\g^{ab}\lb)\*_b
+\eta^{-2}(\lb\g^{ab}\lb)(\lb\g^{cd}r)(\l\g_{bcd}D)\cr
&-16\eta^{-3}(\lb\g^{a[b}\lb)(\lb\g^{cd}r)(\lb\g^{e]f}r)
        (\l\g_{fb}\l)(\l\g_{cde}w)\punkt\cr}\Eqn\ROperator
$$ 
Some alternative ways of writing the last term are given in Appendix A.
Here, the invariant $\eta$ is defined as $\eta=(\l\g^{ab}\l)(\lb\g_{ab}\lb)$

The action of ref. [\PureSG] is
$$
S_3
=\int[dZ]\left[\fr2\Psi Q\Psi+\fr6(\l\g_{ab}\l)\Psi R^a\Psi R^b\Psi\right]
\punkt\Eqn\TwoThreeAction
$$
In order to show that the BV master equation is
fulfilled to third order in the field, we only need to use
$$
(\l\g_{ab}\l)[Q,R^b]=0\komma\Eqn\QRZero
$$
which was how $R^a$ was constructed in
ref. [\PureSG]. (There, $R^a$ was viewed as an operator from the space of
scalar functions $\Psi$ to the space of vectorial functions $\Phi^a$ 
with the extra gauge invariance
$\Phi^a\approx\Phi^a+(\l\g^a\varrho)$. The factor $(\l\g_{ab}\l)$ in
eq. (\QRZero) encodes this invariance.)

\subsection\CSTerm{An example: The Chern--Simons term}In
ref. [\PureSG], it was shown that one of the terms contained in the 
three-point coupling gave the ghost couplings appropriate for the
diffeomorphism algebra. 
We would like to complement that example with one clearly displaying
how a
known interaction among physical supergravity fields, namely the
supergravity Chern--Simons term $\int C\wedge H\wedge H$, is generated.
The cohomology for the $C$-field is [\GrassiVanhove]
$$
\Psi_C\sim(\l\g^i\th)(\l\g^j\th)(\l\g^k\th)C_{ijk}+\ldots\komma\eqn
$$
where the ellipsis
denotes terms with $H=dC$ (which are higher order in $\th$). Acting
with $R_0^a$ gives 
$R_0^a\Psi_C\sim\eta^{-1}
(\lb\g^{ai}\lb)(\l\g^j\th)(\l\g^k\th)(\l\g^l\th)\*_iC_{jkl}
+\ldots$.
The integrand in the coupling term becomes
$$
\eqalign{
&\sim\eta^{-1}(\lb\g^{ij}\lb)
 (\l\g^k\th)(\l\g^l\th)(\l\g^m\th)C_{klm}\cr
&\qquad\times
(\l\g^n\th)(\l\g^p\th)(\l\g^q\th)\*_iC_{npq}
(\l\g^r\th)(\l\g^s\th)(\l\g^t\th)\*_jC_{rst}\punkt\cr
}\Eqn\CSIntOne
$$
We now use the identity [\BerkovitsMembrane]
$(\l\g^{i_1}\th)\ldots(\l\g^{i_9}\th)\sim\e^{i_i\ldots i_9ab}(\l\g_{ab}\l)
{\cal N}$, where ${\cal N}$ is the scalar cohomology at $\l^7\th^9$
used in the measure. We also replace $\*_iC_{jkl}$ by $H_{ijkl}$ in
eq. (\CSIntOne), since the only way to form a scalar is ``$C\wedge
H\wedge H$''. Inserting this in (\CSIntOne) directly gives
$\sim{\cal
  N}\e^{i_i\ldots i_{11}}C_{i_1i_2i_3}H_{i_4i_5i_6i_7}H_{i_8i_9i_{10}i_{11}}$,
without the need of adding any $q$-exact terms. This shows that the
supergravity Chern--Simons term, and hence by supersymmetry all
supergravity 3-point 
couplings, are contained in the interaction term.

\section\CompleteDynamics{The complete dynamics}

\immediatesubsection\TheFullAction{The full action}We will now examine 
the master equation to higher order. 
The master equation reads
$$
(S,S)=0\komma\eqn
$$
where the antibracket is defined as 
$$
(A,B)=\int
A{\larrowover\d\over\d\Psi(Z)}[dZ]{\rarrowover\d\over\d\Psi(Z)}B
\punkt\eqn
$$
We begin by performing a variation of the action (\TwoThreeAction):
$$
\eqalign{
\d S_3&=\int[dZ]\d\Psi\left[Q\Psi+\fr6(\l\g_{ab}\l 
)R^a\Psi R^b\Psi+\fr3R^a\left((\l\g_{ab}\l)\Psi
  R^b\Psi\right)\right]\cr
&=\int[dZ]\d\Psi\left[Q\Psi+\fr2(\l\g_{ab}\l 
)R^a\Psi R^b\Psi+\fr3\Psi R^a\left((\l\g_{ab}\l)
  R^b\Psi\right)\right]\punkt\cr
}\eqn
$$
The remainder from the antibracket 
is
$$
(S_3,S_3)=\fr3\int[dZ]
(\l\g_{ab}\l)R^a\Psi R^b\Psi\Psi R^c((\l\g_{cd}\l)R^d\Psi)
\punkt\Eqn\MasterRemainder
$$
Here, we have already used $(\l\g_{ab}\l)(\l\g_{cd}\l)R^a\Psi R^b\Psi
R^c\Psi R^d\Psi=0$, which follows from the pure spinor constraint.
If the last term would vanish, \ie, if
$R^a(\l\g_{ab}\l)R^b=0$, 
the action $S_3$ would be the full action.
We will now show that this is 
not true, but almost so, in the sense that $R^a((\l\g_{ab}\l)R^b\Psi)$
is a cohomologically trivial field.

The detailed calculation is performed in Appendix B. It is a bit
lengthy, but once it is performed it leads to very simple properties
for the operators. The calculation in Appendix B shows that
$$
R^a(\l\g_{ab}\l)R^b=\fr2(\l\g_{ab}\l)[R^a,R^b]=\Fr32\{Q,T\}
\punkt\eqn
$$
where we have defined 
the fermionic operator $T$ with dimension 3 and ghost number
$-3$ as
$$
T=8\eta^{-3}(\lb\g^{ab}\lb)(\lb r)(rr)N_{ab}\punkt\eqn
$$
Note that $T\Psi$ is bosonic and has dimension as well as ghost number
zero. It seems likely that the field $T\Psi$ is
connected to the trace of the metric fluctuation, and thus to the
determinant of the metric.
So, the operator $R^a(\l\g_{ab}\l)R^b$ is zero in the cohomology.
In addition, the operator $T$ has very nice properties. Since it contains a
multiplicative factor $(\lb r)$, it squares to zero, even when two
$T$'s act on different fields, $TA\,TB=0$. 
Consequently, $TA\{Q,T\}B+\{Q,T\}ATB=0$ (for fermionic $A$), 
and in particular 
 $T\Psi\{Q,T\}\Psi=0$.
$T$ commutes with $R^a$, and of course with the
regularisation factor in the measure (since $R^a$ does). 
It does not commute with
$(\l\g_{ab}\l)$, but as long as there are contraction with $R^a$ and
$R^b$ the commutator gives zero: $R^aAR^bB[T,(\l\g_{ab}\l)]C=0$.

The remaining term in the master equations may now be written
$$
(S_3,S_3)=\fr2\int[dZ]
(\l\g_{ab}\l)\Psi\{Q,T\}\Psi R^a\Psi R^b\Psi
\punkt\eqn
$$
This term is cancelled by the antibracket between a term 
$-\fr4\int[dZ](\l\g_{ab}\l)\Psi T\Psi R^a\Psi R^b\Psi$ and the kinetic
term in the action:
$$
S=\int[dZ]\left[\fr2\Psi Q\Psi+
\fr6(\l\g_{ab}\l)(1-\Fr32T\Psi)\Psi R^a\Psi R^b\Psi\right]\punkt\Eqn\FullAction
$$
With this slight modification of the action given in ref. [\PureSG],
the master equation is exactly satisfied. 
Due to the simple algebraic
properties of $T$, only terms to linear order in $T\Psi$ appear, and
no new terms of higher order in $\Psi$ are generated in $(S,S)$ (see
below for an explicit demonstration of this fact using the equation of
motion). 
Somewhat surprisingly, we thus find that the full supergravity action
in the pure spinor superfield formulation is polynomial.

The factor $1-{3\over2}T\Psi$ in the coupling term may be removed by
performing a field
redefinition $\Psi=(1+{1\over2}T\TPsi)\TPsi$ (which due to the
nilpotency of gives $T\TPsi=T\Psi$ and the inversion
$\TPsi=(1-{1\over2}T\Psi)\Psi$).  
This leads to
a non-canonical kinetic term:
$$
\eqalign{
S
&=\int[dZ]\left[\fr2(1+T\TPsi)\TPsi Q\TPsi+
\fr6(\l\g_{ab}\l)\TPsi R^a\TPsi R^b\TPsi\right]\cr
&=\int[dZ]\left[\fr2e^{T\TPsi}\TPsi Q\TPsi+
\fr6(\l\g_{ab}\l)\TPsi R^a\TPsi R^b\TPsi\right]\punkt\cr
}\Eqn\RedefinedAction
$$
This action is probably closer related to the geometric formulation
than the action (\FullAction).
Note that the field redefinition is not canonical with respect to the
antibracket, 
which can be calculated using 
${\d\over\d\Psi}=(1-T\TPsi){\d\over\d\TPsi}+\fr2\TPsi T{\d\over\d\TPsi}$.

The equation of motion following from the redefined action enjoys
cancellations between terms from the kinetic term and the coupling
term and reads
$$
(1+\Fr32T\TPsi)Q\TPsi+\fr2 (\l\g_{ab}\l)R^a\TPsi R^b\TPsi=0\komma\eqn
$$
while the equation of motion for the canonical field is
$$
Q\Psi+\fr2\Psi\{Q,T\}\Psi+\fr2(\l\g_{ab}\l)(1-2T\Psi)R^a\Psi R^b\Psi=0
\punkt\Eqn\EoM
$$

Using the equation of motion, it is easy to show explicitly that the
master equation is indeed satisfied to all orders. The master equation
is equivalent to the vanishing of the integral of the square of the
equation of motion (for the canonical field). The square of each of
the three terms in eq. (\EoM) gives zero, and the cross terms are
$$
\eqalign{
(S,S)=\int[dZ]\bigl[
Q\Psi\Psi\{Q,T\}\Psi&+(\l\g_{ab}\l)Q\Psi(1-2T\Psi)R^a\Psi R^b\Psi\cr
&+\fr2(\l\g_{ab}\l)\Psi\{Q,T\}\Psi R^a\Psi R^b\Psi
\bigr]\cr}\eqn
$$
Using the algebraic properties of the operators above, it is
straightforward to show that the terms both at third and fourth order
in $\Psi$ combine into total derivatives.

Since the operators involve negative powers of
$\eta=(\l\g_{ab}\l)(\lb\g^{ab}\lb)$,  the singular
properties ar $\eta=0$ must be checked. In ref. [\PureSG], it was
shown that the number of negative powers of $(\l\g^{ab}\l)$ or
$(\lb\g^{ab}\lb)$ must be
smaller than $12$ for an integral to converge. Each $R^a$ has at most
4 negative powers, while $T$ has 5. The expression (\FullAction) needs
regularisation in order to be well defined. This can probably be
achieved using a method similar to that of ref. 
[\BerkovitsNekrasovMultiloop], but it is not
obvious to what extent the algebraic properties of $R^a$ and $T$ will
be preserved by such a regularisation. The form (\RedefinedAction) 
of the action, on
the other hand, is well defined without regularisation.

\subsection\ChangeOfBackground{Expansion around a background}Let
$\Psi_0$ be a solution to the equation of motion (\EoM) and let 
$\Psi=\Psi_0+\psi$. We choose to
expand the ``canonical'' action (\FullAction), 
since the field $\psi$ is canonical
in the sense that the antibracket is
$$
(A,B)=\int
A{\larrowover\d\over\d\psi(Z)}[dZ]{\rarrowover\d\over\d\psi(Z)}B
\komma\eqn
$$
which allows us to compared the expanded and original actions
directly. An expansion of the non-canonical action (\RedefinedAction)
requires letting
$\TPsi=\TPsi_0+(1-{1\over2}T\TPsi_0)\tpsi-\TPsi_0T\tpsi$, but can also
be obtained from rescaling of result below in terms of $\psi$.
Expanding
around the solution gives
$$
S=S[\Psi_0]+\int[dZ]\left[\fr2\psi Q'\psi
+\fr6(\l\g_{ab}\l)(1-\Fr32T\psi)\psi R'^a\psi
R'^b\psi\right]
\komma\Eqn\BackgroundAction
$$
where
$$
\eqalign{
Q'&=Q+2Q\Psi_0T+(1-2T\Psi_0)
\left[(\l\g_{ab}\l)R^a\Psi_0 R^b+\fr2\Psi_0\{Q,T\}\right]\komma\cr
R'^a&=(1-T\Psi_0)R^a-2R^a\Psi_0T\punkt\cr
}\Eqn\NewOperators
$$
It may be checked directly that $Q'^2=0$ when $\Psi_0$
fulfills the equation of motion. The commutators 
$(\l\g_{ab}\l)[Q',R'^b]$ (being $0$ for $\Psi_0=0$)
and $(\l\g_{ab}\l)R'^aR'^b$ (equalling
${3\over2}\{Q,T\}$ for $\Psi_0=0$), seem to become more complicated,
however. Obviously, the master equation will hold, but we have not
checked this explicitly using the primed operators. The master
equation implies relations \eg\
$(\l\g_{ab}\l)[Q',R'^a]\psi R'^b\psi=0$, which are implied by the
previous ones (the ones in a flat background)
but may be weaker in the sense that they need
contractions with fields.

It is nice to verify that there is a kind of weak background
invariance, in the sense that the action in any background is given by
the same formal expression, given by eq. (\BackgroundAction), with
background dependent operators fulfilling the same relations
independent of background (although weaker relations than the ones
used in the flat background). 
Since the model contains gravity, it is natural that the BRST operator
encodes information about the background geometry.
We have not yet been able to examine the connection between the action
in a background and the corresponding construction starting from a
solution in superspace. There, one would construct the BRST operator
as ${\cal Q}=\l^\a D_\a=\l^\a E_\a{}^M\*_M$, $E_A{}^M$ being the inverse
supervielbein of the background. 
It is reasonable to expect a relation (equality?) between the
operators ${\cal Q}$ and $Q'$ of eq. (\NewOperators), and also between
interaction terms. In order to establish such a relation, one should
perform the analogous construction to the one in the present paper and
ref. [\PureSG] but in a curved background superspace. The calculation
of Appendix B relies on the algebra of flat superspace derivatives,
and has to be revised in other backgrounds. We find it likely that the
calculation will stay formally unchanged with the flat derivatives
replaced by covariant derivatives in other backgrounds, but this
remains to be seen.
 
The fact that the action is polynomial of
course gives a small hope of finding a truly background independent
formulation. The solution $\Psi=0$ corresponds to flat space. In a
background independent formulation the expectation value $0$ for the
field would be a non-geometric situation, and flat space would arise
through an expectation value of the field.

\section\Conclusions{Conclusions}Contrary to the expectations
expressed in ref. [\PureSG], the interaction term derived there turned
out to be almost the complete answer. The action for
eleven-dimensional supergravity turns out to be polynomial, and only
contains up to four-point couplings (or three-point, after a field
redefinition). Once the algebraic relations between the operators used
in the construction are derived, the construction encodes the full
nonlinear structure of the supergravity in an extremely simple way.
The efficiency of the pure spinor formalism in reducing the complexity
of supersymmetric dynamics, already demonstrated for $D=10$
super-Yang--Mills theory [\BerkovitsParticle] 
and the BLG and ABJM models in $D=3$ [\CederwallBLG,\CederwallABJM],
turns out to be present also for supergravity. We have no clear
understanding why this happens.

The formulation was made specifically in a flat background, although
it was shown in Section \ChangeOfBackground\ that the action takes the
same formal expression in any background. We have however not yet been
able to relate that action to one obtained from the supergeometry of
the background. 
The geometric status of the supersymmetric action is somewhat
unclear. It should be stressed, though, that the full gauge
invariance, consisting of superdiffeomorphisms and tensor gauge
symmetry (together with an infinite number of cohomologically trivial
symmetries), is present, if not completely covariant.
The precise relation of the pure spinor formulation and the geometric
formulation needs to be clarified. 
Such a relation would hopefully resolve the issue of background invariance.
Maybe some improvement of the pure
spinor action could make it more geometric and simplify the comparison.
One possibility may be the introduction of a field $\Omega^a{}_b$
containing a spin connection, making the Lorentz symmetry local.
On the other hand, it seems to some extent 
to be the ``de-geometrisation'' of the
action that allows for the supersymmetric formulation.

It may not be as strange as it sounds to have a polynomial action for
gravity, once auxiliary fields are included.
Recall the
first order formulation of gravity with an independent spin
connection,
$S\sim\int \e_{a_1\ldots a_d}e^{a_1}\wedge\ldots\wedge
e^{a_{D-2}}\wedge R^{a_{D-1}a_D}(\omega)$. 
The equation of motion from varying the spin connection is the the
torsion-free condition on the vielbein, which eliminates the spin
connection as an independent field. The dynamics becomes non-polynomial
when the torsion constraint is solved, since the solution involves the
inverse vielbein.
Something similar may be happening in the pure spinor formalism. 
There is more than enough room in the superfield at ghost
number 0 to accommodate a spin connection.

The formulation treats metric and tensor degrees of freedom in a
democratic way. 
This may open for a simple proof of U-duality in dimensional
reductions of the model. We envisage two possible ways of dealing with
U-duality.
One possibility is to  try to incorporate the compact subgroup of the U-duality
group as an enlarged structure group, which will involve new types of
pure spinors and new cohomology. Another would be to try to realise
U-duality operators on the field
$\Psi$ as ghost number 0 operators constructed with non-minimal pure
spinors.

The quantum properties of $N=8$ supergravity are not completely
understood. A formulation with manifest supersymmetry would provide a
good starting point. Some calculations have already been made for
$D=11$ supergravity using pure spinors in a superparticle formalism
[\AnguelovaGrassiVanhove], but having a field-theoretic action from
which amplitudes are derived will put the formalism on more solid ground.
In order to use the action for calculating amplitudes one needs to
perform gauge fixing. An essential part of this is to find the
$b$-ghost, with the property $\{Q,b\}=\square$. 
The $b$-ghost for pure spinor superfields in $D=10$ was given in ref.
[\BerkovitsNonMinimal]. It is singular when $(\l\lb)=0$, \ie, at the
tip of the the pure spinor c\^one. The $r$-independent part of that
operator, $b_0\sim(\l\lb)^{-1}(\lb\g^aD)\*_a$ does not work in $D=11$,
in that it does not satisfy $\{s,b_0\}+\{q,b_1\}=0$ for any $b_1$. We
envisage that the singular behaviour instead comes with negative
powers of $\eta=(\l\g_{ab}\l)(\lb\g^{ab}\lb)$, like in the operator
$R^a$. Work on gauge fixing is under way.

Gauge fixing of a BV action amounts to ordinary gauge
fixing of the physical fields together with elimination of the
antifields. A condition $b\Psi=0$ is not the whole story, since one
would like the antifields for the metric and tensor fields not to be
set to zero, but to be related to the corresponding ``antighosts'' in a
(field-theoretically) non-minimal 
BV formalism [\FusterBVReview]. The standard
procedures for gauge fixing (which treat fields and antifields
asymmetrically) are not applicable in a setting where all
fields and antifields reside in the single field $\Psi$. 
It is likely that extra fields need to be introduced, containing
Nakanishi--Lautrup fields and antighosts.
These aspects
have to our knowledge not been addressed for pure spinor superfield
theory, and should be investigated.  

Finally, there will be need to regularise operators which diverge on
some subspace of pure spinor space. We have not dealt with this
problem yet, since at least one of the alternative forms of the action
turned out to be well defined without regularisation. Hopefully,
a method similar to the one in ref. [\BerkovitsNekrasovMultiloop] will work.

\acknowledgements The author would like to thank Nathan Berkovits for
discussions. 


\refout

\vfill\eject

\appendix{Spinor and pure spinor identities in $D=11$}We will list
some identities that have been useful for calculations.

Fierz rearrangements are always made
between spinors at the right and left of two spinor products. The
general Fierz identity reads
$$
(AB)(CD)=\sum\limits_{p=0}^5
\Fr{1}{32\,p!}(C\g^{a_1\ldots a_p}B)(A\g_{a_p\ldots a_1}D)\eqn
$$  
(with appropriate signs for statistics of operators).
For bilinears in a pure spinor $\l$ this reduces to
$$
(A\l)(\l B)=-\fr{64}(\l\g^{ab}\l)(A\g_{ab}B)
+\fr{3840}(\l\g^{abcde}\l)(A\g_{abcde}B)\punkt\eqn
$$ 
From the constraint on the spinor $r$, $(\lb\g^a r)=0$, one derives 
$$
(\lb\g^{[ij}\lb)(\lb\g^{kl]}r)=0\punkt\Eqn\LambdaBarRRelOne
$$
Other useful relations among the non-minimal variables include
$$
\eqalign{
&(\lb\g^i{}_k\lb)(\lb\g^{jk}r)=(\lb\g^{ij}\lb)(\lb r)\komma\cr
&(\lb\g^i{}_kr)(\lb\g^{jk}r)=(\lb\g^{ij}r)(\lb
r)+\fr2(\lb\g^{ij}\lb)(rr)\komma\cr
&(\lb\g^i{}_k\lb)(\lb\g^k{}_lr)(\lb\g^{lj}r)=0\komma\cr
&(\lb\g^i{}_kr)(\lb\g^k{}_lr)(\lb\g^{lj}r)=0\punkt\cr
}\Eqn\LambdaBarRRelTwo
$$
These can used to show quite directly that 
$(\l\g_{ab}\l)[R_1^a,R_1^b]=0$ (see Appendix B).

The symmetry of $(\lb\g^{a[b}\lb)(\lb\g^{cd}r)(\lb\g^{e]f}r)$ is
$(af)$, and no contraction is allowed, so this expression lies in the
irreducible module $(20010)$.


Various useful identities for a pure spinor $\l$ include
$$
\eqalign{ 
&(\g_j\l)_\a(\l\g^{ij}\l)=0\komma\cr
&(\g_i\l)_\a(\l\g^{abcdi}\l)=6(\g^{[ab}\l)_\a(\l\g^{cd]}\l)\komma\cr
&(\g_{ij}\l)_\a(\l\g^{abcij}\l)=-18(\g^{[a}\l)_\a(\l\g^{bc]}\l)\komma\cr
&(\g_{ijk}\l)_\a(\l\g^{abijk}\l)=-42\l_\a(\l\g^{ab}\l)\komma\cr
&(\g_{ij}\l)_\a(\l\g^{abcdij}\l)=-24(\g^{[ab}\l)_\a(\l\g^{cd]}\l)\komma\cr
&(\g_i\l)_\a(\l\g^{abcdei}\l)=\l_\a(\l\g^{abcde}\l)
-10(\g^{[abc}\l)_\a(\l\g^{de]}\l)\komma\cr}\eqn
$$

Alternative forms for $R_2^a$ are:
$$
\eqalign{
R_2^a&=-16\eta^{-3}(\lb\g^{a[b}\lb)(\lb\g^{cd}r)(\lb\g^{e]f}r)
(\l\g_{fb}\l)(\l\g_{cde}w)\cr
&=\Fr43\eta^{-3}(\lb\g^{ab}\lb)(\lb\g^{cd}r)(\lb\g^{ef}r)(\l\g_{bcde}{}^g\l)N_{fg}
\cr
&\qquad-\Fr23\eta^{-3}(\lb\g^{ab}\lb)(\lb\g^{cd}r)(\lb
r)(\l\g_{bcd}{}^{ef}\l)N_{ef}\cr
&=2\eta^{-3}\left[\eta(\lb\g^{ab}r)-2\phi(\lb\g^{ab}\lb)\right]
(\lb\g^{cd}r)(\l\g_{bcd}w)\cr
&=\{s,\eta^{-2}(\lb\g^{ab}\lb)(\lb\g^{cd}r)\}(\l\g_{bcd}w)\komma\cr
}\eqn
$$
where $N_{ab}=(\l\g_{ab}w)$ and $\phi=(\lb\g^{ij}r)(\l\g_{ij}\l)$.


\appendix{Calculation of a commutator}In this Appendix, we will
calculate the operator $R^a(\l\g_{ab}\l)R^b$ appearing in the master
equation after the three-point coupling is introduced, and thus
governing higher interactions. We can write
$R^a(\l\g_{ab}\l)R^b=[R^a,(\l\g_{ab}\l)]R^b+\fr2(\l\g_{ab}\l)[R^a,R^b]$.

Consider the first term. The
only non-vanishing contribution comes from $R_2^a$. Using the form
from Appendix A,
$R_2^a=2\eta^{-3}\left[\eta(\lb\g^{ab}r)-2\phi(\lb\g^{ab}\lb)\right]
(\lb\g^{cd}r)(\l\g_{bcd}w)$, 
one gets
$$
[R^a,(\l\g^{bc}\l)]
=4\eta^{-3}\left[\eta(\lb\g^{ai}r)-2\phi(\lb\g^{ai}\lb)\right]
(\lb\g^{jk}r)(\l\g_{ijkbc}\l)
\punkt\Eqn\RLL
$$ 
If two of the indices are contracted, this gives zero thanks to
$(\lb\g^{[ij}\lb)(\lb\g^{kl]}r)=0$.

The second term takes some more work. Examine first the terms in
$[R^a,R^b]$ coming from the commutator of $w$ with one of the
prefactors $\eta^{-k}$. Using eq. (\RLL) again, we get
$$
[R^a,\eta]
=-4\left[\eta(\lb\g^{ai}r)-2\phi(\lb\g^{ai}\lb)\right]
(\lb\g^{jk}r)(\l\g_{ijkbc}\l)(\lb\g^{bc}\lb)=0\punkt\eqn
$$
Now, the only remaining things to check are the terms from the
anticommutator of the two $D$'s in $R_1$ and from the commutator of
the $w$ in $R_2$ with $\l$'s in $R_1$ and $R_2$ (except in $\eta$).
Anticommuting the two $D$'s in $R_1$ gives
$$
(\l\g_{ab}\l)[R_1^a,R_1^b]
=\eta^{-3}(\lb\g^{ai}\lb)(\lb\g^{bc}r)(\lb\g^{jk}r)
(\l\g_{abc}\g^m\g_{ijk}\l)\*_m\punkt\Eqn\RTwoRTwo
$$
Expanding the product of $\g$-matrices, 
$$
(\l\g_{abc}\g^m\g^{ijk}\l)\*_m
=3(\l\g_{abc}{}^{[ij}\l)\*^{k]}+3(\l\g_{[ab}{}^{ijk}\l)\*_{c]}
-9\d^{[i}_{[a}(\l\g_{bc]}{}^{jk]m}\l)\*_m\punkt\eqn
$$ 
The corresponding three
terms in eq. (\RTwoRTwo) vanish individually due to the identities 
(\LambdaBarRRelOne) and (\LambdaBarRRelTwo) in
Appendix A.

Similarly, the commutator between $R_1$ and $R_2$ gives
$$
\eqalign{
2(\l\g_{ab}\l)[R_2^a,R_1^b]
&=4\eta^{-5}(\l\g_{ai}\l)\left[\eta(\lb\g^{ab}r)-2\phi(\lb\g^{ab}\lb)\right]
(\lb\g^{cd}r)\cr
&\qquad\times(\lb\g^{ij}\lb)(\lb\g^{kl}r)(\l\g_{bcd}\g_{jkl}D)\punkt\cr
}\eqn
$$
Here, we encounter the first non-vanishing contribution.
Again using the relations from Appendix A, 
the second
term in the square brackets vanishes. In the first term, there is the
possibility to contract two indices between two separate pairs of matrices
$(\l\g^{ab}\l)$ or $(\l\g^{ab}r)$, thus avoiding the zeroes of the
last two identities in eq. (\LambdaBarRRelTwo). The result is
$$
2(\l\g_{ab}\l)[R_2^a,R_1^b]=24\eta^{-3}(\lb\g^{ab}\lb)(\lb r)(rr)(\l\g_{ab}D)
\punkt\eqn
$$
(From BRST invariance, it also necessary that the part 
with the smallest number of $r$'s does not contain $(\l\g^{(4)}D)$ or
$(\l\g^{(6)}D)$.)
Finally, if we write $R_1^a=M^{abcd}(\l\g_{bcd}D)$ and
$R_2^a=\{s,M^{abcd}\}(\l\g_{bcd}w)$, the above result may be written
$$
\eqalign{
2(\l\g_{ab}\l)[R_2^a,R_1^b]&=2(\l\g_{ab}\l)\{s,M^{aijk}\}M^{blmn}
                     (\l\g_{ijk}\g_{lmn}D)\cr
          &=[q,24\eta^{-3}(\lb\g^{ab}\lb)(\lb r)(rr)(\l\g_{ab}w)]\komma\cr
(\l\g_{ab}\l)[R_2^a,R_2^b]&=2(\l\g_{ab}\l)\{s,M^{aijk}\}\{s,M^{blmn}\}
                     (\l\g_{ijk}\g_{lmn}w)\cr
          &=[s,24\eta^{-3}(\lb\g^{ab}\lb)(\lb r)(rr)(\l\g_{ab}w)]
          \komma\cr
}\eqn
$$
and thus
$$
(\l\g_{ab}\l)[R^a,R^b]=[Q,24\eta^{-3}(\lb\g^{ab}\lb)(\lb
r)(rr)(\l\g_{ab}w)]\punkt\eqn
$$

\end